# Comparing Topology of Engineered and Natural Drainage Networks

Soohyun Yang[1,2], Kyungrock Paik[2*], Gavan McGrath[1,3], Christian Urich[4],
Elisabeth Kruger[1,5], Praveen Kumar[6], and P. Suresh C. Rao[1,7]

[1]Lyles School of Civil Engineering, Purdue University, West Lafayette, IN, USA
[2]School of Civil, Environmental, and Architectural Engineering,
Korea University, Seoul, South Korea
[3]Ishka Solutions, Perth, Western Australia, Australia
[4]Civil Engineering Department, Monash University, Melbourne, Victoria, Australia
[5]Helmholtz Centre for Environmental Research - UFZ, Leipzig, Germany
[6]Department of Civil and Environmental Engineering,
University of Illinois at Urbana-Champaign, Urbana, IL, USA
[7]Agronomy Department, Purdue University, West Lafayette, IN, USA

*Corresponding author: Kyungrock Paik (paik@korea.ac.kr)

**Abstract**

We investigated the scaling and topology of engineered urban drainage networks (UDNs) in two cities, and further examined UDN evolution over decades. UDN scaling was analyzed using two power-law characteristics widely employed for river networks: (1) Hack's law of length ($L$)-area ($A$) scaling [$L \sim A^h$], and (2) exceedance probability distribution of upstream contributing area ($\delta$) [$P(A \geq \delta) \sim a\delta^{-\varepsilon}$]. For the smallest UDNs (< 2 km$^2$), length-area scales linearly ($h \sim 1$), but power-law scaling emerges as the UDNs grow. While $P(A \geq \delta)$ plots for river networks are abruptly truncated, those for UDNs display exponential tempering [$P(A \geq \delta) = a\delta^{-\varepsilon} \exp(-c\delta)$]. The tempering parameter $c$ decreases as the UDNs grow, implying that the distribution evolves in time to resemble those for river networks. However, the power-law exponent $\varepsilon$ for large UDNs tends to be slightly larger than the range reported for river networks. Differences in generative processes and engineering design constraints contribute to observed differences in the evolution of UDNs and river networks, including subnet heterogeneity and non-random branching.

# 1 Introduction

Efficient drainage of urban landscapes is among the critical services provided to the citizens for avoiding flooding of streets and neighborhoods, and for maintaining flows to wastewater treatment plants (WWTPs). Here, of particular interest are urban drainage networks (UDNs) which include storm-water, sanitary sewers, and combined sanitary storm-water systems. UDNs are located below-ground at shallow depths and in close proximity to road networks [*Blumensaat et al.*, 2012; *Klinkhamer et al.*, 2017; *Mair et al.*, 2012; *Mair et al.*, 2017]. Our understanding about the network characteristics of engineered UDNs is limited with surprisingly little literature (e.g., *Oh* [2010]). An intriguing question in this regard is how the topology of UDNs compares with their natural analogs, i.e., river networks. UDNs, like rivers, involve gravity-driven and directed flows from the entire drainage area converging towards a WWTP. Many large cities have multiple outlets (e.g., combined sewer overflow outlets; several WWTPs) forming multiple sewer-sheds, whose boundaries may overlap several natural watersheds. UDNs consist of junctions and conduits, which correspond to nodes and edges, similar to confluences and reaches in river networks. Like stream orders in river networks, UDNs exhibit hierarchy of pipe-diameters for a range of designed maximum flows, and flows are directed towards an outlet. These structural and functional similarities prompted the application of river network hierarchical organization concepts to describe UDNs. *Cantone and Schmidt* [2011a, 2011b], *Sitzenfrei et al.* [2013], and *Urich et al.* [2010] classified the hierarchy of real/virtual sewer systems through Horton-Strahler ordering scheme [*Strahler*, 1957]. While river networks evolve through natural processes, UDNs are engineered networks designed to meet efficiently urban drainage requirements at the minimum cost. River networks drain large landscapes, up to the continental scales (~$10^6$ km$^2$; e.g., Amazon; Congo; Nile; Rio de la Plata; Mississippi). In contrast, even the largest known UDNs drain much smaller urban sewer-sheds ($\leq 10^3$ km$^2$) [*USEPA*, 2001].

Analogies and differences (see Table 1) lead us to explore differences in network topology of UDNs to those reported for river networks. It is well-established that river networks are fractal with self-similarity revealed through quantitative scaling relationships [*Hack*, 1957; *Horton*, 1945; *Marani et al.*, 1994; *Rigon et al.*, 1996; *Rodríguez-Iturbe et al.*, 1992a; *Tarboton et al.*, 1991; *Tokunaga*, 1978]. We examine here whether UDNs and river networks share scaling properties reported for river networks, given similarities in their functions in landscape drainage, and despite differences between engineered versus natural networks.

We investigate here the functional organization and scaling of UDNs in terms of their topological features, and examine how scaling patterns change as they grow over several decades. We analyze UDN evolution towards some attractor ("mature" network) based on time-stamped data. We begin with an exploration of scaling relationships of the most recent UDNs, and compare our findings to well-known and universal relationships established for river networks. Then, we examine how the scaling of an engineered drainage network has evolved over time by examining UDN topology during the preceding decades.

*Table 1. Key differences between river networks and urban drainage networks*

| Factors | River Networks | Urban Drainage Networks |
|---|---|---|
| **Generative Processes** | Formed from geomorphological processes, including erosion, weathering, tectonic uplift, ecosystem functions, operating as a driver of natural landscape evolution, over geologic time-scales. (E.g., *Parker* [1977]) | Planned, designed, built, maintained and operated by engineers to drain urban landscapes, and evolves over several decades or even centuries. |
| **Network Size & Flow Direction** | River networks can drain continental-scale basins (~$10^6$ km$^2$), with nested sub-basins ranging from $10^2$ to $10^5$ km$^2$. Rivers carve out flow paths along the steepest downstream gradient. | Largest urban drainage networks are ~$10^3$ km$^2$, usually draining to a single outlet (wastewater treatment plant, WWTP). Large cities (>$10^6$ population) have multiple WWTPs. Sewer lines are often laid out to follow roads rather than in the direction of steepest descent [*Blumensaat et al.*, 2012; *Haghighi*, 2013]. |
| **Direction of Network Growth** | For a given drainage area with an established outlet, starting with the main channel, river networks grow upstream, eventually occupying entire drainage area [*Parker*, 1977]. | The drainage area grows in time as the city grows, and its maximum discharge increases. The initial UDN draining to a WWTP is expanded as new neighborhoods are added to the city. |
| **Network Structure** | Bifurcating, branching trees. Loops occur only under exceptional conditions such as river deltas and braided rivers (p. 231 in *Knighton* [1998]). Drainage area is constrained to the DEM-defined catchment, except under exceptional conditions (e.g., divide migration). | Imperfect branching trees, with 2 or more pipes connected to some junctions. Loops represent less than 1% of network. Drainage areas can cross natural watershed boundaries. When loops represent a significant fraction of UDNs, methods have been proposed for defining equivalent binary-tree networks for looped configurations [*Haghighi*, 2013; *Seo and Schmidt*, 2014]. |
| **Optimality** | For example, minimize total energy dissipation [*Paik and Kumar*, 2010; *Rodríguez-Iturbe et al.*, 1992b], given a fixed drainage area delineated by topography (DEM), under the assumption of uniform forcing (constant rainfall) over a heterogeneous landscape. | Maximum engineering efficiency for designed total discharge (at the outlet), and minimize costs. Most projects pursue local optimizations in time and space, often constrained by funds. Expansion of the system after repeated local optimization may not result in optimality achieved over time. |

## 2 Study areas and data analyses

Detailed data of UDNs are difficult to obtain, and recent emphasis on security concerns and privatization of utilities have further limited data accessibility; thus, our initial analysis is restricted to a few UDNs for which data recorded over several decades were available. We examined two UDNs (Wahiawa and Honouliuli) from Oahu Island, Hawaii, USA (~1 million population), and the largest UDN from a large anonymous Asian city (AAC) (~4 million population). These cities have contrasting socio-economic backgrounds, history, climate, and terrain. The Oahu UDN data are publicly available while the AAC data were obtained under a confidentiality agreement from the city's water utility authority. Node-degree distribution and other topological metrics for AAC, based on dual mapping, were presented in *Krueger et al.* [2017]. Here, we analyzed the AAC's largest sewer-shed (drainage area: ~100 km$^2$; population ~2.4 million). A summary of the characteristics of each network is provided in Table 2.

To calculate upstream drainage area for every node in a UDN, the open-source platform *DynaMind* (http://iut-ibk.github.io/DynaMind-ToolBox/) was used [*Urich et al.*, 2012]. We retrieved data for UDNs at five-year intervals utilizing pipe installation records. Beginning at a WWTP, from its construction date, a continuous pipe network was determined by incrementally sampling upslope pipes with the same or earlier installation date although not all pipes contained attribute data for installation date.

*Table 2*: *Characteristics of the UDNs at the last year of record.*

|  | **Wahiawa** | **Honouliuli** | **AAC** |
|---|---|---|---|
| **Data period** | 1929-2009 | 1923*-2014 | 1968-2015 |
| **Drained area (km$^2$)** | 7.2 | 85.1 | 126.0 |
| **Number of pipes** | 1,364 | 17,255 | 49,355 |
| **Length of pipes (m)** | 77,725 | 929,205 | 1,895,459 |
| **Number of loops** | 8 | 23 | 252 |

\* *Website for Hawaii Water Environment Associate (http://www.hwea.org/) states that the Honouliuli WWTP was put into service in December 1984.*

Two relationships, widely found in river networks, were used in this study to address whether UDNs exhibit scaling behavior is comparable to that in rivers. The first, Hack's Law, is a power-law relationship between main channel length, *L*, and its corresponding drainage area *A*:

$$L = A^h \qquad (1)$$

where the exponent *h* is reported to be universally in a small range, $h = 0.6\pm0.1$ [*Crave and Davy*, 1997; *Hack*, 1957; *Paik and Kumar*, 2011; *Robert and Roy*, 1990]. The other scaling relationship is the exceedance probability distribution of the upstream drainage area [*Rodríguez-Iturbe et al.*, 1992a] within catchments retrieved from digital elevation models (DEMs). For rivers, the exceedance probability that the upstream drainage area (*A*) is equal to or greater than a value $\delta$ is reported to follow a power-law. For UDNs, this can be expressed as an exponentially tempered Pareto distribution [*Aban et al.*, 2006],

$$P(A \geq \delta) = a\delta^{-\varepsilon} \exp(-c\delta); \quad \delta \geq \delta_{min} \qquad (2)$$

The exponent $\varepsilon$ for most river networks is 0.43±0.03 (and $c = 0$, representing abrupt truncation) suggesting universality of this scaling relationship [*Crave and Davy*, 1997; *Maritan et al.*, 1996; *Rinaldo et al.*, 2014; *Rodríguez-Iturbe et al.*, 1992a].

## 3 Results and discussion

### 3.1 Scaling Patterns for "Quasi-Mature" UDNs

Plots of $P(A \geq \delta)$ for the most recent UDNs exhibit similar patterns (Figure 1a); a straight middle part, here called "trunk," and the tempered "tail" resulting from the finite-size effect. The trunk section becomes more evident for larger UDNs (e.g., AAC). Maximum drainage areas ($A_{max}$) for these UDNs (< 130 km$^2$) are much smaller than typical river networks. UDNs are constrained by the city size and economy-of-scale constraints on the size of the treatment plants; often, multiple UDNs drain large cities. Notably for UDNs, finite-size constraints are shown as the smooth exponential truncation of $P(A \geq \delta)$. In contrast, finite-size effects for river networks are evident as power-law truncation at all spatial scales [*Rinaldo et al.*, 2014].

From Eq. (2), $P(A \geq \delta)$ can be normalized as $P(A \geq \delta)(a\delta^{-\varepsilon})^{-1}$ [*Rinaldo et al.*, 2014]. Our normalized exceedance probability distribution collapse onto a single curve, albeit with some variation toward the end of the tempered tails (Figure 1b). This suggests comparable topologies among the UDNs regardless of diverse (e.g., climatic, demographic, economic, engineering, and geographic) constraints underpinning the evolution of these UDNs.

For contributing area $\geq 0.02$ km$^2$, we estimated both $c$ and $\varepsilon$ in Eq. (2) for each UDN. Fitting Eq. (2) involves three degrees of freedom: (1) the lower threshold where the power-law begins; (2) the upper threshold where the power-law diminishes; and (3) the power-law exponent $\varepsilon$. The upper threshold is parameterized as $c$ in Eq. (2). For contributing area equal to and greater than the lower threshold, the two parameters ($\varepsilon$ and $c$) in Eq. (2) were found using Matlab's *nlinfit* function of which the objective function is to minimize the sum of the squares of the residuals for the fitted model. The estimated ranges for each parameter were calculated with 95% confidence intervals.

We also investigated the underlying river networks of our study area. Based on DEMs, we extracted river networks for catchments corresponding to the urban areas of our study. We fitted the exceedance probability (Eq. (2)) without exponential tempering, i.e., $P(A \geq \delta) = a\delta^{-\varepsilon}$ as is widely done for river networks. Estimated $\varepsilon$ values for these natural drainage networks for $\delta > \delta_{min}$ (=0.01 km$^2$) are, as expected, in the range reported for other river networks, i.e., 0.43±0.03 [*Crave and Davy*, 1997; *Maritan et al.*, 1996; *Rinaldo et al.*, 2014; *Rodríguez-Iturbe et al.*, 1992a]. In comparison, the overall mean $\varepsilon$ value of UDNs depends on their size. We find $\varepsilon$ values of 0.36, 0.46, and 0.53, respectively, for Wahiawa (year 2009; 126 km$^2$), Honouliuli (year 2014; 85.1 km$^2$), and AAC (year 2015, 7.2 km$^2$). A small UDN like Wahiawa shows a strong exponential tempering with a short power-law range fitted with a smaller $\varepsilon$ value. A large UDN of AAC, on the other hand, exhibits a long power-law range with sharp tail truncation, comparable with river networks. This implies that UDN has the potential to become mature in the future, exhibiting a robust power-law, as its drainage area expands (subject to size constraints as discussed above). For such a large UDN for AAC, $\varepsilon$ (0.53), is greater than that reported for rivers. This can be partly because UDNs are not strict binary trees, unlike river networks, and can have more than two upstream pipes draining into some junctions. We estimated such junctions to be ~22% for Wahiawa, ~19% for Honouliuli, and ~19% for AAC UDNs.

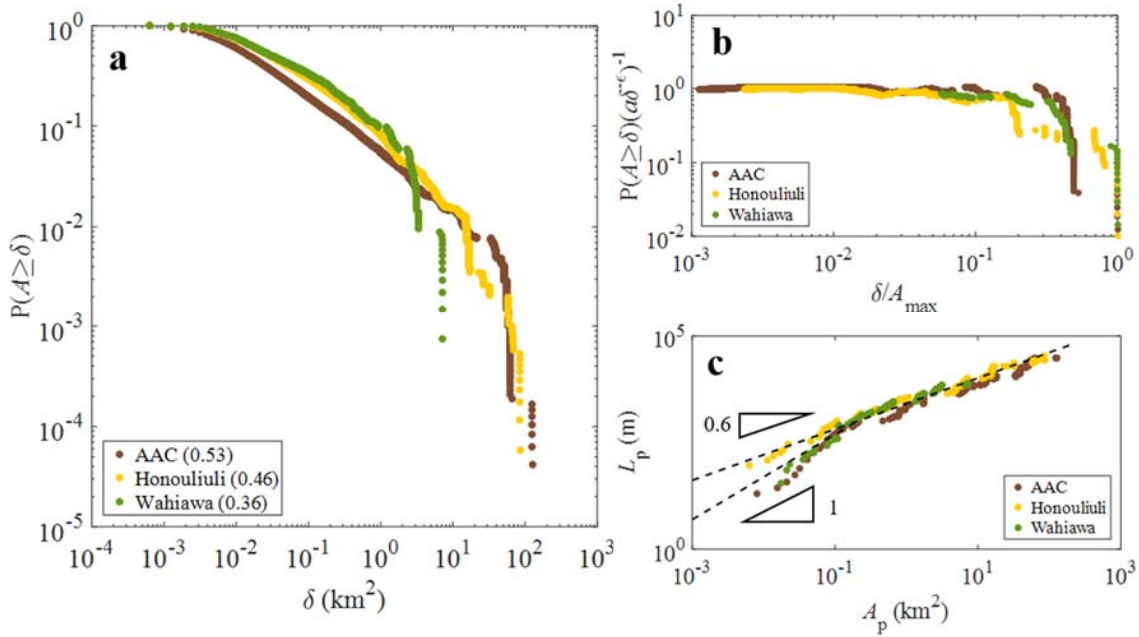

*Figure 1.* Scaling relationships of three UDNs of most recent data. (*a*) Area-exceedance probability distributions for AAC, Honouliuli, and Wahiawa networks (total drainage area $A_{max}$ are 126.0 km$^2$, 85.1 km$^2$, and 7.2 km$^2$, respectively). Numbers in the parentheses of legend are the mean fitted $\varepsilon$ value in Eq. (2). Standard errors for the $\varepsilon$ value with 95% confidence interval are: 0.001 (AAC), 0.002 (Honouliuli), and 0.005 (Wahiawa). Mean squared error values for the fitted model are: $3.3\times10^{-6}$ (AAC), $3.1\times10^{-5}$ (Honouliuli), and $7.0\times10^{-5}$ (Wahiawa). (*b*) Normalized area-exceedance probability distribution with power-law fit. (*c*) Area-length relationships for the three UDNs. $L_p$ is the pipe length along main pipe line and $A_p$ is the drainage areas corresponding to $L_p$. The two dotted black lines are displayed as the "ideal slopes" for comparison with the data (linearity h=1; Hack's law exponent h=0.6).

The relationship between pipe length and area also follows a power-law (Figure 1c), like Hack's law in river networks. We postulate the following scenario for the emergence of power-law scaling with the growth of UDNs. Given an initial single sewer line, the length-area relationship of such a simple case is linear and thus $h \approx 1$. As the UDN grows, repeated additions of other pipes along its length from adjoining branches lead to a self-repetitive tree, but still far from self-similarity and thus having a linear length-area relationship. Nevertheless, with continuous network growth, some random factors such as the size of added subnets and junction locations are introduced, which is sufficient to drive the transition from this deterministic, repetitive structure, to that approaching a self-similar tree [*Paik and Kumar*, 2007].

We observe that length-area relationships are often dissected into two segments, i.e., upstream segment, which shows the convergence to $h \approx 1$, and the downstream segment with $h \approx 0.6$. The threshold between these segments is about 0.2 km$^2$, much greater than $\delta_{min}$ in Figure 1a. This implies that the topological reasoning for the threshold in length-area relationship differs from the reasoning for $\delta_{min}$. The downstream segment is considered as the 'mature' section of a UDN where enough number of branches have been connected. The upstream segment is the terminal section, corresponding to the new development area at the outskirt where early settlement forms only along the single major line which is connected to the main body of the network, without development of background blocks. As cities expand such areas are densified, but another new development forms further away from the city core. Hence, such break point in the length-area relationship may occur at any stage of a growing city. The break point, however, is invisible in Figure 1a because $P(A \geq \delta)$ is for the entire network rather than a single path.

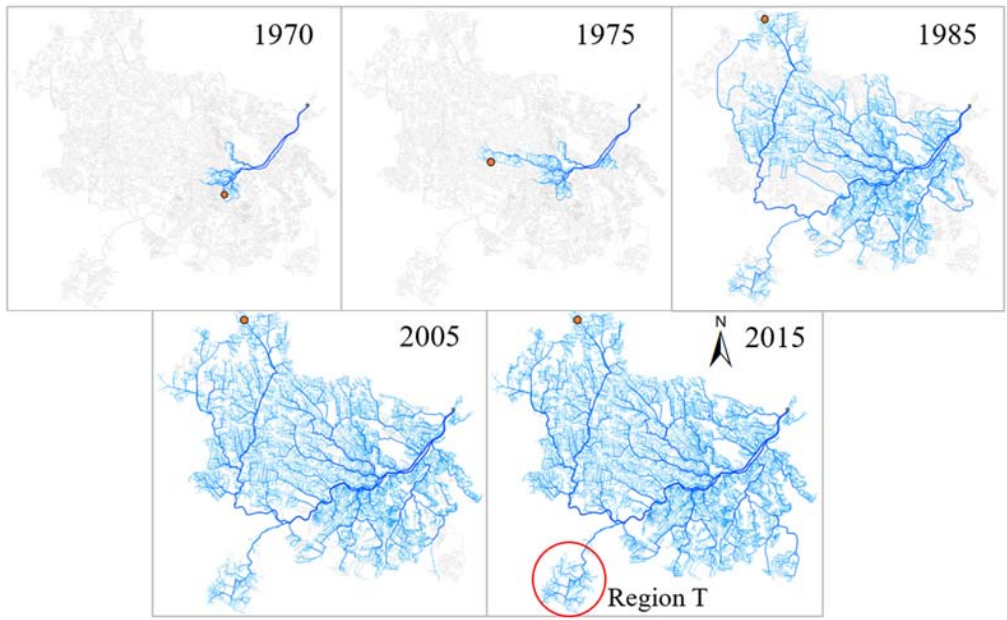

*Figure 2. Evolution of the pipe layout (blue lines) for the UDN in AAC over decades (1970-2015). Line thickness is proportional to the Horton-Strahler order of each pipe branch. The latest UDN is shown in the background with grey color. An orange dot represents the node of which upslope area is 0.2* km$^2$*, on the main branch (the visible break point in Hack's law in Figure 1c). Detailed network configuration, upstream from the orange dot is investigated [data not shown].*

Scaling constants, $\varepsilon$ and $h$, are related as $\varepsilon + h \approx 1$ for natural river networks [*Rigon et al.*, 1996]. Given estimated values of $\varepsilon$ and $h$ in Figure 1, such constraint is not evident for UDNs. This is because the space-filling constraint in river network is not relevant for UDNs. For example, a single sewer-line, which connects a sub-sewershed (e.g., a satellite Region T in Figure 2) to the main network, can be lengthy but has a little lateral contributing area. Therefore, unlike river networks, decoupling between $\varepsilon$ and $h$ is likely for UDNs.

The long-term expansion of sewer networks occurs in an episodic fashion, with short periods of quick growth, either to accommodate anticipated growth in demand, or to connect new neighborhoods with increasing population, or because of retrofitting catalyzed by technological advancements. Such spurts in UDN growth are reflected largely in step-like nature in the tails of the area exceedance plots (Figure 1b). In river networks, similar dynamic shifts have been shown to occur, but at orders-of-magnitude less frequent rates over geologic times, as revealed in stream piracy or migration of drainage divides [*Bonnet*, 2009; *Willett et al.*, 2014] resulting from tectonic shifts or non-stationary climate forcing. UDN expansion can also involve pipe networks crossing topographic watershed divides of naturally drained catchments.

### 3.2 Scaling of Evolving UDNs

Water infrastructure networks in cities evolve over time involving two parallel processes: (1) *expansion,* with increase in area with the addition of new suburbs; and (2) *densification,* when older neighborhoods become filled-in or expanded vertically [*Gudmundsson and Mohajeri*, 2013; *Mohajeri et al.*, 2015]. Both processes contribute to an increase in total pipe length and number of network nodes in proportion to population [*Krueger et al.*, 2017; *Zischg et al.*, 2017].

We investigated the evolution of the pipe layout for the AAC UDN shown in Figure 2, and the associated scaling relationships over several decades (Figure 3). In the scaling relations for evolving UDN, $\varepsilon$ values estimated for the most recent networks well represent the "trunk" portion of the area exceedance plots (Figures 3a, c, e). Similarly, the length-area scaling (Figures 3b, d, f) for evolving UDNs mirrors that seen in their more mature state (Figure 1c).

The power-law exponent $\varepsilon$ in Eq. (2) increases as the total drainage area, $A_{max}$, increases over time (Figure 3g). On the other hand, the exponential tempering parameter, $c$, scales inversely with increasing $A_{max}$ (Figure 3h), with an apparent log $c$ versus log $A_{max}$ slope of 0.73. These results suggest that as UDNs grow to drain larger urban areas, the exponential tempering diminishes, with $P(A \geq \delta)$ tending to be more abruptly truncated, thus increasingly resembling river networks.

As a city grows, drainage density tends to increase in a given area of a city. Figure 4a shows an example of growing UDN sub-nets within a 53 km$^2$ boundary area of AAC. The growing UDN within the bounded region shows well-collapsed $P(A \geq \delta)$ curves (Figure 4b). Our repeated analysis for total four sub-networks in AAC shows similar evolving patterns, with elongation of the trunk and more abrupt truncation over time [data not shown].

The total length of pipes within each clipped area of the UDN increases rapidly during the initial decade or two, representing the period of expansion, followed by much slower increase over the next decades (Figure 4c), consistent with empirical analyses of city growth [*Gudmundsson and Mohajeri*, 2013; *Mohajeri et al.*, 2015]. Exponential tempering of $P(A \geq \delta)$ diminishes as the sub-networks grow within the clipped area, with $c$ decreasing inversely with increasing total pipe length (Figure 4d). Thus, $c$ is a measure of the stages and patterns of UDN expansion and densification. The rate at which $c$ decreases varies among different subnets in the city (drainage areas ranging from 7 to 46 km$^2$), reveals the heterogeneity of UDN growth in AAC.

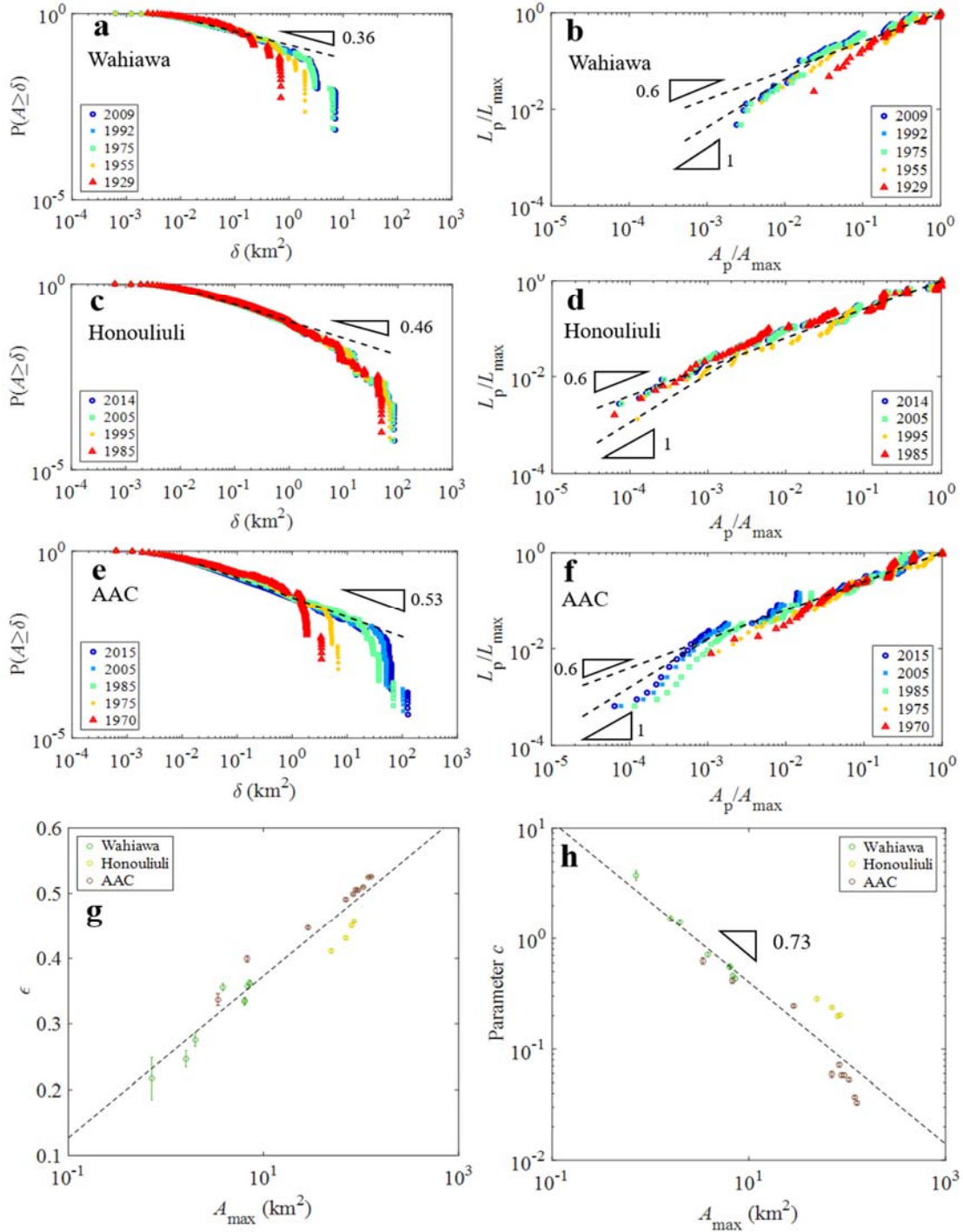

*Figure 3.* Scaling relationships of three evolving UDNs: Wahiawa (1929-2009); Honouliuli (1985-2014); and AAC (1970-2015). (***a, c, e***) Area-exceedance probability distribution plots. (***b, d, f***) Power-law relationship between normalized length ($L_p/L_{max}$, where $L_{max}$ is a maximum length of main pipeline in each year) - normalized area ($A_p/A_{max}$). The two dotted black lines are shown as the "ideal slope" (linearity h=1 and Hack's law exponent h=0.6) for comparison. (***g***) Trend of the exponent $\varepsilon$ in Eq. (2) over varying maximum sewer-shed drainage area (Black dashed line is fitted as $\varepsilon = 0.053 \ln A_{max} + 0.25$). (***h***) Power-law relationship between the maximum sewer-shed drainage area and the truncation parameter c (fitted power-law exponent -0.73 with standard error 0.13 for 95% confidence interval).

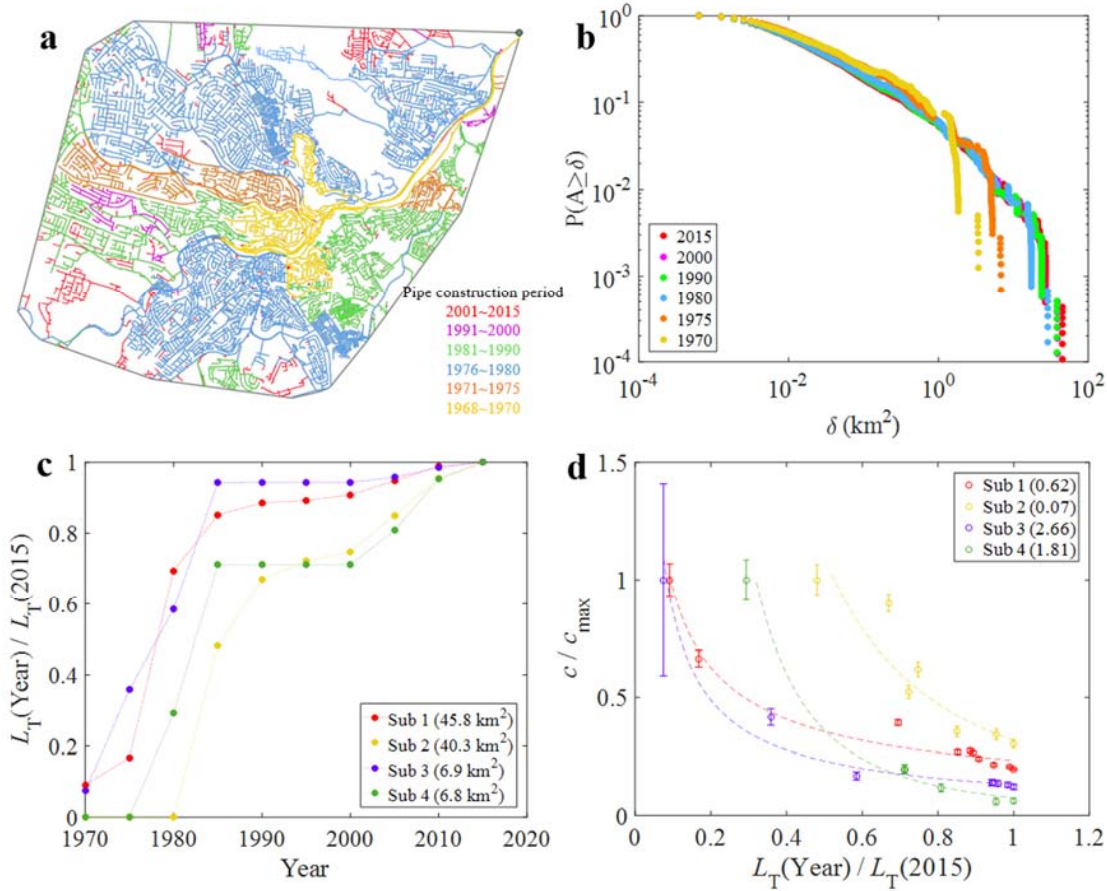

*Figure 4. Evolution of sub-networks in AAC. (**a**) Map of the sub-network 1, which is within a clipped area (53 km$^2$) of AAC. (**b**) Area-exceedance probability distribution of growing pipes shown in Figure 4a. (**c**) Growing total pipe length over time. Total pipe length in a certain year $L_T$(Year) is normalized by the longest total pipe length in the latest $L_T$(2015). Numbers in the parenthesis of legend are the maximum drainage area in 2015. Sub 1 indicates the sub-network shown in Figure 4a. [Other sub-networks data are not shown]. (**d**) Variation of exponential tempering parameter c, normalized to the maximum c value at the earliest time $c_{max}$. Numbers in the parenthesis of legend are $c_{max}$ values. 95% confidence intervals are also shown as the bars.*

## 4 Conclusions

### 4.1. UDN Topology

1. Abrupt truncation is evident in $P(A \geq \delta)$ for rivers, while UDN exhibit exponential tempering, with the tempering constant $c$ decreasing with the UDN size. The relationship between pipe length and drainage area also follows a power-law, like Hack's law for river networks, dissected into two segments: an upstream segment with $h \approx 1$, and a downstream segment with $h \approx 0.6$. Further, $\varepsilon + h \approx 1$ for natural river networks, but not for UDNs because the space-filling constraint in river network is not relevant for UDNs. These results suggest that with growing size, UDN topology increasingly resembles that of river networks, with abrupt truncation of power-law $P(A \geq \delta)$ from finite-size effects.

2. The power-law exponent $\varepsilon$ for a UDN depends on the network size and can be larger than that reported for river networks. This suggests that these engineered networks have a larger hierarchical density, expected of an imperfect non-random branching tree topology.

Multiple factors controlling the range of $\varepsilon$ for UDNs should be explored using data from diverse cities.
3. Different UDN subnets within a city evolve through both expansion and densification, but at heterogeneous growth rates, reflected in decreasing rate of the tempering constant $c$ with growth. Observed variability among UDN subnets within a city and among UDNs for different cities is reflective of engineering constraints that generate an imperfect, non-random tree structure.

**4.2. UDN Evolution**

Adjustments or additions to growing UDN necessitate adaptive engineering design improvements in order to maintain functionality throughout a UDN. Such incremental engineering practices at local- and city-scale have strong parallels to "self-organization," where the city and its infrastructure networks continually evolve, and are adjusted to meet expansion and changing needs. Indeed, several studies have shown that cities evolve to exhibit fractal geometries, with optimal space-filling physical networks and assets [*Batty and Longley*, 1994; *Batty et al.*, 1989; *Shen*, 2002]. *Lu and Tang* [2004] showed that as cities grow to occupy increasingly larger areas, the urban spaces are filled more densely by city roads, which are increasingly more fractal, providing more efficient access to all locations within the city. Geospatial co-location of roads and sewers suggests similar functional topology for UDNs and roads. The importance of "local" engineering optimization evident in small subnets of UDNs diminishes as the "global" optimization becomes more dominant at larger scales. Thus, large UDNs appear to be "self-organized" though differently to rivers. Further analysis of other UDNs from diverse cities is required to support or contradict these conclusions; such efforts are underway.

**Acknowledgments**

This study is an outcome from the Synthesis Workshop on "*Dynamics of Structure and Functions of Complex Networks*", held at Korea University in 2015. This work was jointly supported by the U.S. NSF Award Number 1441188 (Collaborative Research-RIPS Type 2: Resilience Simulation for Water, Power and Road Networks) and National Research Foundation of Korea (NRF) grant funded by the Korean government (MSIP) (grant number 2015R1A2A2A05001592). The authors also acknowledge insightful comments offered by Dr. Andrea Rinaldo during the early phases of the development of concepts presented here.